\documentclass[aps,pra,reprint,superscriptaddress,floatfix]{revtex4-1}
\usepackage{graphicx}
\usepackage{amsmath,amssymb,cases,bm} 

\begin{document}

\title{Correspondence between a shaken honeycomb lattice and the Haldane model}

\author{Michele Modugno}
\affiliation{\mbox{Depto. de F\'isica Te\'orica e Hist. de la Ciencia, Universidad del Pais Vasco UPV/EHU, 48080 Bilbao, Spain}}
\affiliation{IKERBASQUE, Basque Foundation for Science, 48011 Bilbao, Spain}

\author{Giulio Pettini}
\affiliation{Dipartimento di Fisica e Astronomia, Universit\`a di Firenze,
and INFN, 50019 Sesto Fiorentino, Italy}

\date{\today}

\begin{abstract}
We investigate the correspondence between the tight-binding Floquet Hamiltonian of a periodically modulated honeycomb lattice and the Haldane model. We show that -- though the two systems share the same topological phase diagram, as reported in a breakthrough experiment with ultracold atoms in a stretched honeycomb lattice [Jotzu \textit{et al.}, Nature \textbf{515}, 237 (2014)] -- the corresponding Hamiltonians are not equivalent, the one of the shaken lattice presenting a much richer structure. 
\end {abstract}

\maketitle

\section{Introduction}

The Haldane model \cite{haldane1988} is a paradigmatic lattice model characterized by a quantum Hall effect due to the breaking of time-reversal symmetry in the presence of a microscopic magnetic field with vanishing flux across the unit cell. The distinctive feature of its phase diagram is a phase transition between normal and topological insulating phases, depending on the value of the phase $\varphi$ of the next-nearest tunneling amplitude. Despite the fact that it cannot be directly implemented in conventional condensed matter systems, the Haldane model has a fundamental importance as the mechanism for nontrivial band topology can be realized in actual materials via the intrinsic spin-orbit interaction of topological insulators \cite{hasan2010,qi2011}. 

Very recently, novel opportunities for investigating topological phases have been opened by means of periodically driven optical lattices, which indeed represent an extraordinary platform for simulating a wide class of Hamiltonians \cite{eckardt2005, kitagawa2010,hauke2012,delplace2013,jotzu2014,goldman2014,baur2014,verdeny2015,plekhanov2017}.
In particular, a realization of the Haldane topological phases has been reported in a breakthrough experiment with ultracold atoms in a \textit{stretched} honeycomb lattice \cite{jotzu2014}, whose position is moved along a closed elliptical path, $\bm{r}_{lat}(t)=-A(\cos(\omega t)\bm{e}_{1}+ \cos(\omega t-\theta)\bm{e}_{2})$. 
In this experiment, the Haldane topological phase diagram can be exactly reproduced when drawn as a function of the phase $\theta$ of the modulation, that experimentally is the parameter directly associated to the breaking of time-reversal symmetry, instead of the phase of the next-nearest tunneling coefficients as in the original Haldane model. 
Actually, though it has been proven that an imaginary component is generated in the next-nearest tunneling coefficients of the effective Hamiltonian at the lowest orders of the Magnus expansion (powers of $1/\omega$) \cite{jotzu2014,verdeny2015}, a formal correspondence with the Hamiltonian of the Haldane model is still lacking.

Motivated by these considerations, in this paper we consider the case of a \textit{regular} honeycomb lattice 
\cite{kitagawa2010,ibanez-azpiroz2013,verdeny2015}, periodically modulated along the same path of Ref. \cite{jotzu2014}, and we discuss the structure of the corresponding tight binding model and its relation with the Haldane model. We show that the effective Hamiltonian of the shaken lattice has a very rich structure and cannot be directly identified with the Haldane model. In fact, we find that there is no formal correspondence between the phase $\theta$ of the modulation and the phases $\varphi_{ij}$ of the next-nearest tunneling amplitudes. This result has no implications for the topological phase diagram, that is actually very robust and does not depend on the details of the model, but just on the local structure at the Dirac points \cite{montambaux2009,montambaux2009b,fuchs2012,tarruell2012,ibanez-azpiroz2013b,delplace2013}. Rather, it clarifies that -- just thanks to this robustness -- a tight binding model defined on a honeycomb lattice and characterized by topological phases in the presence of parity and time reversal symmetry breaking, is not necessarily described by the Hamiltonian of the Haldane model \cite{bukov2015}. In fact, as shown in Ref. \cite{verdeny2015}, the exact simulation of this model requires an optimal design of the lattice driving, by means of polychromatic forces. In a certain sense, the Haldane topological phase diagram identifies a \textit{universality class} of Hamiltonians sharing the same topological behavior.

The paper is organized as follows. In Sect. \ref{sec:honeycomb} we review the properties of a regular honeycomb lattice and of the Haldane model. Then, in Sect. 
\ref{sec:modulation} we discuss the periodic modulation, and the corresponding transformation that maps the honeycomb Hamiltonian in the rotating frame, following the protocol of Ref. \cite{jotzu2014}. In Sect. \ref{sec:effective} we define the effective Floquet Hamiltonian in momentum space and discuss its numerical implementation. There we also discuss the corresponding topological phase diagram, and the structure of the resulting tight binding model on the direct lattice. A discussion of the general properties of the model, and of its behavior in the limits of fast modulations and of small amplitudes (in momentum space), is presented in Sect. \ref{sec:discussion}. Finally, concluding remarks are drawn in Sect. \ref{sec:conclusions}.

\section{Honeycomb lattice and Haldane model}
\label{sec:honeycomb}

\begin{figure}[t]
\centerline{\includegraphics[width=\columnwidth]{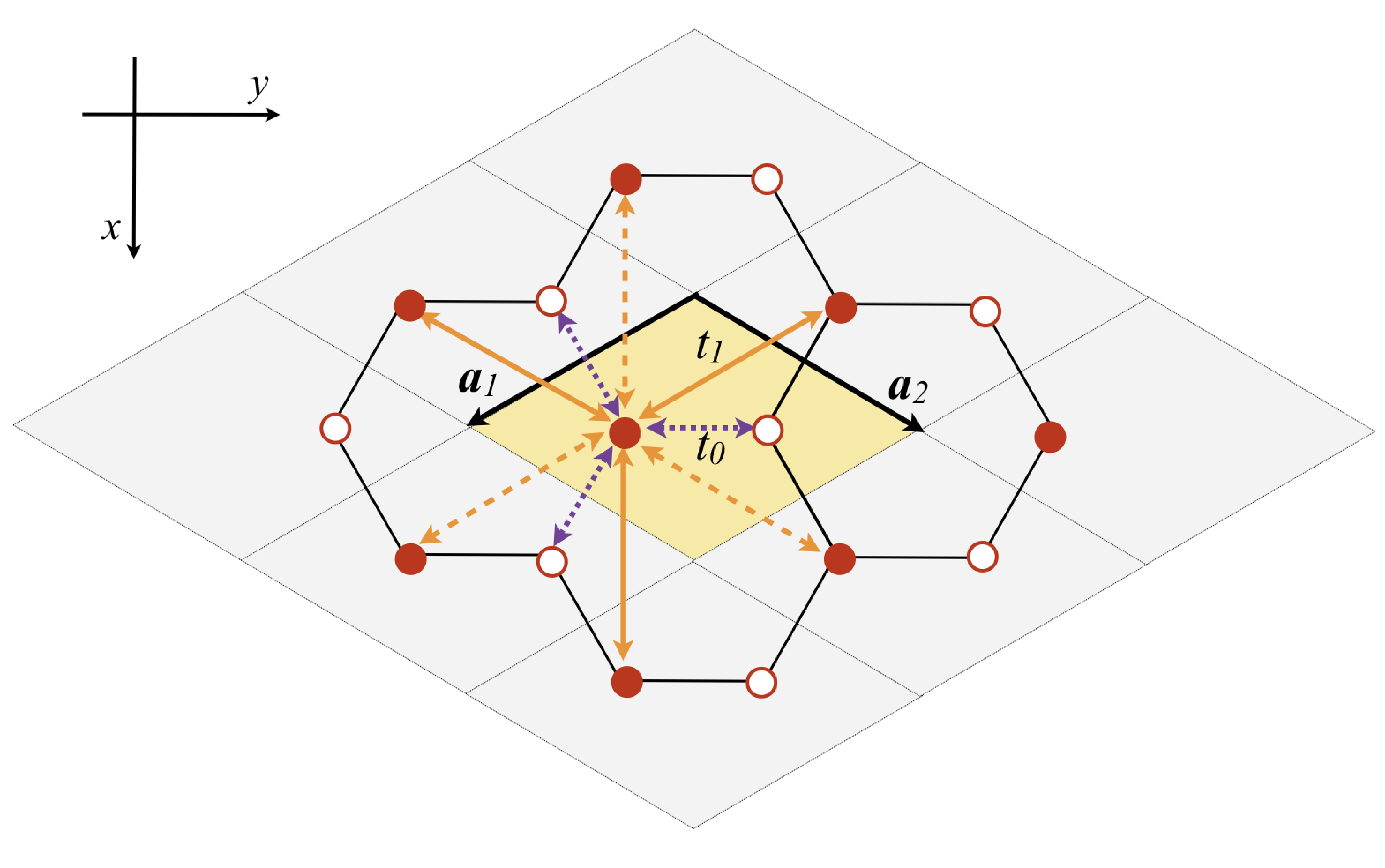}}
\caption{Sketch of the honeycomb lattice. Filled and empty circles refer to minima of type $A$ and $B$, respectively. The elementary cell is highlighted in yellow. The tunneling coefficients up to next-nearest neighbors ($t_{0}$, $t_{1}$) are indicated for the site of type $A$ in the central cell. The length of each side of the hexagon is $a=4\pi/(3\sqrt{3}k_{L})$.}
\label{fig:honeycomb}
\end{figure}

Let us consider a honeycomb lattice generated by the vectors 
$\bm{a}_{1/2} =({2\pi}/{3k_{L}}) (\bm{e}_{x}\mp\sqrt{3}\bm{e}_{y})$
as in Fig. \ref{fig:honeycomb}, with basis points $A$ and $B$.
The corresponding tight-binding model up to next-nearest neighbors reads \cite{lee2009}
\begin{equation}
H=-\sum_{\langle ij\rangle}t_{0}\hat{c}_{i}^{\dagger}\hat{c}_{j}
-\sum_{\langle\langle ij\rangle\rangle}t_{1}\hat{c}_{i}^{\dagger}\hat{c}_{j}+\sum_{i}\sum_{\nu=A,B}\epsilon_{\nu}\hat{c}_{i}^{\dagger}\hat{c}_{i}
\label{eq:honeytb}
\end{equation}
where the operators $\hat{c}_{j}$ may be fermionic or bosonic (the following discussion is completely independent of the nature of the particles), and the tunneling amplitudes $t_{0}$ and $t_{1}$ can be chosen real 
\cite{ibanez-azpiroz2013}. Here the explicit parity breaking introduced by an energy offset $\epsilon_{A}-\epsilon_{B}$ between sites of type $A$ and $B$ is assumed to be negligible at the level of the next-nearest tunneling amplitudes, namely $t_{1}^{A}=t_{1}^{B}=t_{1}$ (this is usually the case, see e.g.\cite{jotzu2014,ibanez-azpiroz2015}). 

In momentum space (see Appendix), the Hamiltonian (\ref{eq:honeytb}) 
is represented by the following $2\times2$ matrix \cite{modugno2016,ibanez-azpiroz2013}

\begin{equation}
h(\bm{k})=\left(\begin{array}{cc}
 \epsilon_{A}(\bm{k}) + t_{1}F(\bm{k}) & t_{0}Z(\bm{k}) \\
 t_{0}Z^{*}(\bm{k}) & \epsilon_{B}(\bm{k}) + t_{1}F(\bm{k})
\end{array}\right)
\end{equation}
with
\begin{align}
Z(\bm{k})&=1 +e^{i\bm{k}\cdot\bm{a}_{1}}
+e^{-i\bm{k}\cdot\bm{a}_{2}},
\\
F(\bm{k})&=2\sum_{i=1}^{3}\cos\left(\bm{k}\cdot\bm{a}_{i}\right),
\end{align}
with $\bm{a}_{3}\equiv\bm{a}_{1}+\bm{a}_{2}$.
It can also be rewritten in a compact form, by using the basis formed by the $2\times 2$ identity matrix, $I$, and of the three Pauli matrices, $\sigma_{i}$. One has 
\begin{equation}
 h(\bm{k})=h_{0}(\bm{k})I+\bm{h}(\bm{k})\cdot\bm{\sigma},
 \label{eq:paulidecomp}
\end{equation}
with ${\bm{h}}\equiv(h_{1},h_{2},h_{3})$ and 
\begin{align}
h_{0}(\bm{k})&= t_{1}F(\bm{k}),
\label{eq:h0}\\
h_{1}(\bm{k})&= t_{0}{\rm{Re}}[Z(\bm{k})]
,\label{eq:h1}\\
h_{2}(\bm{k})&= -t_{0}{\rm{Im}}[Z(\bm{k})]
,
\label{eq:h2}\\
h_{3}(\bm{k})&= \epsilon,
\label{eq:h3}
\end{align}
with $\epsilon\equiv(\epsilon_{A}-\epsilon_{B})/2$, and where we have fixed $\epsilon_{A}+\epsilon_{B}=0$ without loss of generality. The corresponding eigenergies are
\begin{align}
\epsilon_{\pm}(\bm{k})&= h_{0}(\bm{k})\pm|{\bm{h}}(\bm{k})|
\nonumber\\
&=t_{1}F(\bm{k})\pm \sqrt{\epsilon^{2} + t_{0}^{2}|Z(\bm{k})|^{2}}.
\label{eq:tbenergies}
\end{align}
In particular, for $\epsilon=0$ the two bands are degenerate at the Dirac points, $\bm{k}_{D}$ where $Z(\bm{k}_{D})=0$.

\subsection{Haldane model}

In the case of the Haldane model, the tunneling coefficient $t_{1}$ becomes complex \cite{haldane1988, ibanez-azpiroz2014,jotzu2014,ibanez-azpiroz2015},
\begin{equation}
t_{1}= |t_{1}|e^{\pm i\varphi},
\label{eq:t1}
\end{equation}
where the sign $\pm$ of the phase $\varphi$ refers to the $A$ and $B$ sublattices, respectively \cite{haldane1988, ibanez-azpiroz2014, ibanez-azpiroz2015}. In momentum space, the components $h_{0}$ and $h_{3}$ of the Hamiltonian, in Eqs. (\ref{eq:h0}), (\ref{eq:h3}), are modified as follows \cite{haldane1988}
\begin{eqnarray}
h_{0}(\bm{k})&=&2|t_{1}|\cos\varphi\sum_{i=1}^{3}\cos\left(\bm{k}\cdot\bm{a}_{i}\right),
\\
h_{3}(\bm{k})&=&\epsilon-2|t_{1}|\sin\varphi\sum_{i=1}^{3}
\sin\left(\bm{k}\cdot\bm{a}_{i}\right).
\end{eqnarray}

In general, when time-reversal and/or inversion symmetry are broken 
there are two inequivalent Dirac points $\bm{k^{\pm}}_{D}$ where a gap opens in the spectrum, namely \cite{modugno2016}
\begin{equation}
 \delta_{\pm}= 2|h_{3}(\bm{k}^{\pm}_{D})|=2\left|\epsilon\pm 3\sqrt{3}|t_1|\sin\varphi\right|.
 \label{eq:gaphaldane}
 \end{equation}
For certain values of $\epsilon$ and $\varphi$ one of the two gaps may close again if ${\bm h}(\bm{k}_{D}^{\pm};\epsilon,\varphi)=0$ (see eq. (\ref{eq:tbenergies})). This relation identifies the boundary between the normal and topological insulator phases \cite{haldane1988}.

As shown in Ref. \cite{ibanez-azpiroz2015,modugno2016}, it is possible to derive a closed set of analytical relations for expressing the tight-binding parameters in terms of specific properties of the spectrum, namely the gaps at the Dirac points $\delta_{\pm}$ in eq. (\ref{eq:gaphaldane}), and the following bandwidths \cite{ibanez-azpiroz2015}
\begin{align}
\Delta^{\pm}_{+}&=+[\epsilon_{+}(\bm{0})-\epsilon_{+}(\bm{k}^{\pm}_{D})],\\
\Delta^{\pm}_{-}&=-[\epsilon_{-}(\bm{0})-\epsilon_{-}(\bm{k}^{\pm}_{D})].
\label{eq:bandwidths}
\end{align}
For $\epsilon,\varphi\geq 0$ one has\footnote{The solutions corresponding to a different regime can be obtained straightforwardly from symmetry considerations, by exchanging the role of the two basis points $A,B$ and/or of the two Dirac points.}
\begin{align}
\label{eq:eps-spc}
\epsilon&=\frac{\delta_{+}\pm\delta_{-}}{4},\\
\label{eq:t0-spc}
t_{0}&=\frac{1}{6}\sqrt{\left(\Delta_+^++\Delta_-^++\delta_{+}\right)^2-\frac{\left(\delta_{+}\pm\delta_{-}\right)^2}{4}},
\\
\label{eq:t1-spc}
|t_{1}|&=\frac{1}{18}\sqrt{\left(\Delta_+^+-\Delta_-^+\right)^2+\frac{3}{4}\left(\delta_{+}\mp\delta_{-}\right)^2},\\
\label{eq:phi-spc}
\tan\varphi&=\left(\frac{\sqrt{3}}{2}\frac{\delta_{+}\mp\delta_{-}}{\Delta_+^+-\Delta_-^+}\right).
\end{align}
where the signs $\pm$ refer to the normal and topological insulator phases, respectively.

In the rest of this paper we analyze the correspondence between a \textit{shaken} honeycomb lattice, that is a Hamiltonian as in (\ref{eq:honeytb}) moved periodically along a closed path, and the Haldane model.

\section{Periodic modulation}
\label{sec:modulation}

Let us consider the effect of a periodic modulation along a closed trajectory $\bm{r}_{lat}(t)=-(q_{0}/m\omega)(\cos(\omega t)\bm{e}_{1}+ \cos(\omega t-\theta)\bm{e}_{2})$ of the bare honeycomb lattice in (\ref{eq:honeytb}) as considered in Ref. \cite{jotzu2014}. 
Classically, in the lattice frame the atoms feel an inertial force $\bm{F}(t)=-m\ddot{\bm{r}}_{lat}(t)\equiv\dot{\bm{q}}_{lat}(t)$, corresponding to a potential term $V(\bm{r},t)=\bm{F}\cdot\bm{r}$, that in the tight binding formalism reads
\begin{equation}
H_{mod}=\sum_{\nu=A,B}\sum_{j}V(\bm{r}_{j\nu},t)\hat{c}_{j\nu}^{\dagger}\hat{c}_{j\nu}.
\end{equation}
 This term can be reabsorbed in the definition of the tunneling coefficients, by means of the following unitary ($\hat{U}^{\dagger}=\hat{U}^{-1}$) transformation acting on the quantum states
\begin{equation}
\hat{U}(t)\equiv e^{ -i\sum_{j,\nu} m\dot{\bm{r}}_{lat}(t)\cdot\bm{r}_{j\nu}\hat{n}_{j}},
\end{equation}
that transforms a lattice Hamiltonian of the form
\begin{equation}
H=-\sum_{\nu\nu'=A,B}\sum_{jj'}t_{jj'}^{\nu\nu'}\hat{c}_{j\nu}^{\dagger}\hat{c}_{j'\nu'} + H_{mod}
\end{equation}
into (rotating frame)
\begin{align}
H^{'}_{lat}&\equiv UHU^{\dagger} + i\dot{U}U^{\dagger}
\\
\nonumber
&=-\sum_{\nu\nu'=A,B}\sum_{jj'}t_{jj'}^{\nu\nu'}e^{ -i m\dot{\bm{r}}_{lat}(t)\cdot(\bm{r}_{j\nu}-\bm{r}_{j'\nu'})}\hat{c}_{j\nu}^{\dagger}\hat{c}_{j'\nu'}.
\end{align}

Then, in the case of the honeycomb Hamiltonian (\ref{eq:honeytb}) we have ($\bm{r}_{AB}=(\bm{a}_{1}-\bm{a}_{2})/3$)
 \begin{align}
Z(\bm{k})&\rightarrow 
e^{ i\bm{q}_{lat}(t)\cdot\bm{r}_{AB}}Z(\bm{k}-\bm{q}_{lat}(t))
\\
F(\bm{k})&\rightarrow F(\bm{k}-\bm{q}_{lat}(t)).
\end{align}
and 
\begin{align}
h_{0}(\bm{k})&= t_{1}F(\bm{k}-\bm{q}_{lat}(t)),
\label{generalhmatrixh0}\\
h_{1}(\bm{k})&= t_{0}{\rm{Re}}[e^{i\bm{q}_{lat}(t)\cdot\bm{r}_{AB}}Z(\bm{k}-\bm{q}_{lat}(t))]
,\label{generalhmatrixh1}\\
h_{2}(\bm{k})&= -t_{0}{\rm{Im}}[e^{i\bm{q}_{lat}(t)\cdot\bm{r}_{AB}}Z(\bm{k}-\bm{q}_{lat}(t))]
,
\label{generalhmatrixh2}\\
h_{3}(\bm{k})&= \epsilon.
\label{generalhmatrixh3}
\end{align}

\section{Effective Hamiltonian}
\label{sec:effective}

For convenience, in the following we will employ the dimensionless time variable $\tau=\omega t$.
Following the discussion in \cite {jotzu2014} (see the Supplementary Material), if one is interested in the dynamics on time scales much larger than one period $\omega T=2\pi$, one can introduce an effective Hamiltonian in terms of the evolution operator $U(\tau+2\pi,\tau)$, as follows \cite{kitagawa2010,shirley1965,bukov2015}
 \begin{equation}
U(\tau+2\pi,\tau)={\cal T}e^{-\displaystyle \frac{i}{\hbar\omega}
\int_{\tau}^{\tau+2\pi}\!\!\!\!\!\!\!\!\!\!\!\!\!d\tau H(\tau)}\equiv e^{\displaystyle-i2\pi h_{eff}^{\tau}},
\label{eq:evolution}
\end{equation}
where ${\cal T}$ indicates the time-ordered product, and $h_{eff}^{\tau}$ the effective Hamiltonian in units of $\hbar\omega$. The expression (\ref{eq:evolution}) has to be intended in the sense of its Taylor expansion, namely
\begin{align}
&{\cal T}e^{-\displaystyle\frac{i}{\hbar\omega}
\int_{\tau}^{\tau+2\pi}\!\!\!\!\!\!\!\!\!\!\!\!\!d\tau H(\tau)}\equiv 
1+ 
\\
\nonumber
&\qquad\sum_{n=1}^{+\infty} \frac{(-i)^{n}}{n!}\frac{1}{\omega^{n}}\int d\tau_{1}\cdots \int d\tau_{n} {\cal T}\left[H(\tau_{1})\cdots H(\tau_{n})\right],
\end{align}
that can also be regarded as an expansion in series of $1/\omega$, known in the literature as Magnus expansion \cite{magnus1954,bukov2015}.
Then, the effective Hamiltonian can be obtained from the inverse relation
\begin{equation}
h_{eff}^{\tau}=i\frac{1}{2\pi}\ln(U(\tau+2\pi,\tau)),
\end{equation}
and it will be calculated numerically as explained in the next section.
Remarkably, the energy spectrum of the effective Hamiltonian does not depend on the choice of the initial time $\tau$, since two effective Hamiltonians for different initial times are related through a unitary gauge transformation, $h_{eff}^{\tau'}=U(\tau', \tau)h_{eff}^{\tau}U^{-1}(\tau',\tau)$. In the following we consider the gauge corresponding to $\tau=0$, a common choice for sinusoidal driving \cite{jotzu2014,bukov2015}. This choice may affect the specific values of the effective tunneling coefficients, but not the general structure of the Hamiltonian.

\subsection{Numerical implementation}

The effective Hamiltonian in momentum space $h_{eff}(\bm{k})$ can be computed numerically from the time-ordered product of $N$ evolution operators $U_{i}=e^{{-i\Delta h(\bm{k},\tau_{i})}}$ over infinitesimal time intervals, as
\begin{equation}
h_{eff}(\bm{k})
=\frac{i}{2\pi}\ln\left(\prod_{i=0}^{N-1}e^{-i\Delta h(\bm{k},\tau_{i})}\right)_{\cal T}
\label{eq:heff}
\end{equation}
with $\Delta\equiv 2\pi/N$, and $N$ sufficiently large. The previous expression can be easily handled in the basis of the Pauli matrices $(I, \bm{\sigma})$, where two matrices $u=(u_{0}, \bm{u})$ and $\eta=(\eta_{0}, \bm{\eta})$, such that $u=e^{-i\eta}$, are related by the following expressions
\begin{equation}
u=e^{-i\eta_{0}}\left(\cos|\bm{\eta}|, -i\frac{\bm{\eta}}{|\bm{\eta}|}\sin|\bm{\eta}|\right)
\label{eq:exp2}
\end{equation}
and 
\begin{equation}
\eta=\left(-\arg(u_{0}),ie^{i\eta_{0}}\frac{|\bm{\eta}|}{\sin|\bm{\eta}|}\bm{u}\right),
\end{equation}
with $|\bm{\eta}|=\cos^{-1}|u_{0}|$.

In the following, we consider a specific implementation with ultracold atoms in optical lattices \cite{lee2009,modugno2016} and use natural lattice scales, that is lengths in units of $k_{L}^{-1}$ ($k_{L}$ being the lattice wave vector) and energies in units of the corresponding recoil energy $E_{R}={\hbar^{2}k_{L}^{2}}/{2m}$. Then, for a regular honeycomb lattice, the tunneling coefficients depend on the lattice amplitude $s$ as \cite{ibanez-azpiroz2013}
\begin{align}
t_{0}/E_R&= 1.16 s^{0.95} e^{-1.634\sqrt{s}},
\label{eq:t0}
\\
t_{1}/E_R &= 0.78 s^{1.85} e^{-3.404\sqrt{s}}
\end{align}
Here we consider a typical tight-binding regime, $s=10$. 
As for the lattice momentum $\bm{q}_{lat}$, we have
\begin{equation}
\bm{q}_{lat}(\tau)
= \frac12q_{0}\left[\sin(\tau)\bm{e}_{1}+ \sin(\tau-\theta)\bm{e}_{2}\right],
\end{equation}
where the correspondence with the notations of Jotzu \textit{et al.} \cite{jotzu2014} is $q_{0}=A\omega$, with 
$A=0.087\lambda\simeq 0.55~k_{L}^{-1}$ and $\omega=2\pi\times4$kHz $\simeq 0.9{E_{R}}/{\hbar}$.

\subsection{Topological phase diagram}

Let us first show that the effective Hamiltonian (\ref{eq:heff}) is characterized by the same topological phase diagram -- as a function of $(\theta,\epsilon)$ -- of the Haldane model as a function of $(\varphi,\epsilon)$, as discussed in Ref. \cite{jotzu2014} for the case of a stretched honeycomb. 
\begin{figure}
\centerline{\includegraphics[width=\columnwidth]{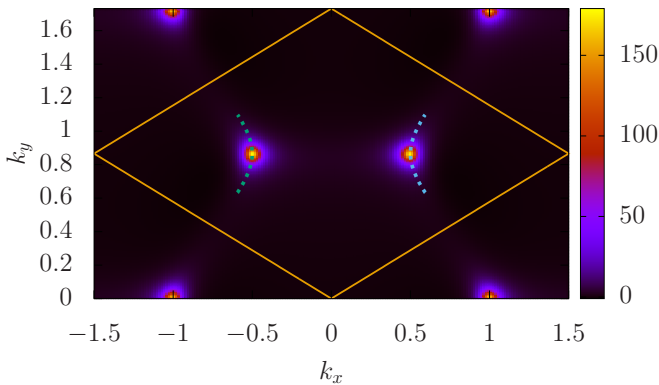}}
\caption{Density plot of the Berry curvature $\Omega(\bm{k})$ within a rhomboidal Brillouin zone containing the Dirac points $\bm{k}_{D}^{\pm}=(\pm1/2,\sqrt{3}/2)k_{L}$, for $\omega=1$, $q_{0}=0.5$, $\theta=\pi/2$, $\epsilon=0$. The two dotted lines represents the trajectory along which the Dirac points move when $\theta$ is varied from $0$ to $\pi$ (upwards). }
\label{fig:bz}
\end{figure}
The topological phase diagram is drawn according to the value of the Chern number $\nu$ of the lowest band. In general, the Chern number for a given band $n$ can be written as
\begin{equation}
c_{n}=\frac{1}{2\pi}\int_{BZ}d^{2}\bm{k}~\Omega_{n}(\bm{k}),
\end{equation}
where $\Omega_{n}(\bm{k})$ is the corresponding Berry curvature
\begin{equation}
\Omega_{n}(\bm{k})= 2{\rm Im}\!\!\sum_{m\neq n}\left[\frac{\langle n_{\bm{k}}| h_{x}(\bm{k})|m_{\bm{k}}\rangle
 \langle m_{\bm{k}}| h_{y}(\bm{k})|n_{\bm{k}}\rangle}
 {(\epsilon_{m}(\bm{k})-\epsilon_{n}(\bm{k}))^2}\right]
\end{equation}
with $|n_{\bm{k}}\rangle$ being the eigenvectors of $h(\bm{k})$, $\epsilon_{n}(\bm{k})$ the corresponding eigenvalues, and $h_{\alpha}(\bm{k})\equiv\partial_{k_{\alpha}} h(\bm{k})$ ($\alpha=x,y$).
In the case of a two band problem, as the present case, we can set $n=1$, $m=2$, and $c=c_{1}$. Since the main contribution to the integral comes from the vicinity of the Dirac points, where the Berry curvature $\Omega(\bm{k})$ is peaked, it is convenient to tile the reciprocal space with rhomboidal cells (like in the coordinate space) containing the two inequivalent Dirac points, see in Fig. \ref{fig:bz}.

\begin{figure}[t!]
\centerline{\includegraphics[width=\columnwidth]{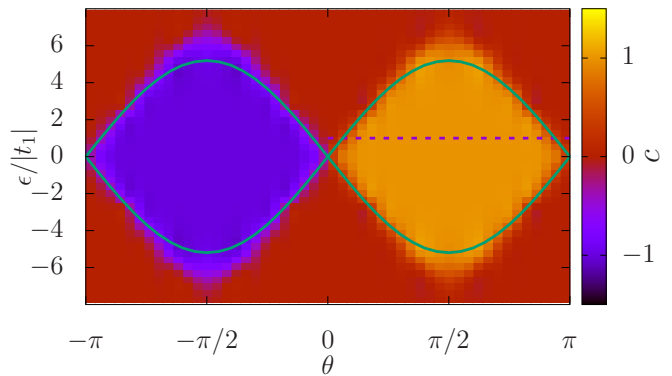}}
\caption{Topological phase diagram of the Hamiltonian (\ref{eq:heff}), for $\omega=1$, $q_{0}=0.5$. The color palette refers to the values of Chern number $c$. The continuous line represents the nominal phase boundary of the Haldane model with $t_{1}'=t_{1}$ and $\varphi=\theta$, see Eq. (\ref{eq:gaphaldane2}), that is shown here as a reference. The dashed line represents a cut along which we draw the gap at the Dirac points in Fig. \ref{fig:gap}.}
\label{fig:pd}
\end{figure}

\begin{figure}[b]
\centerline{\includegraphics[width=0.95\columnwidth]{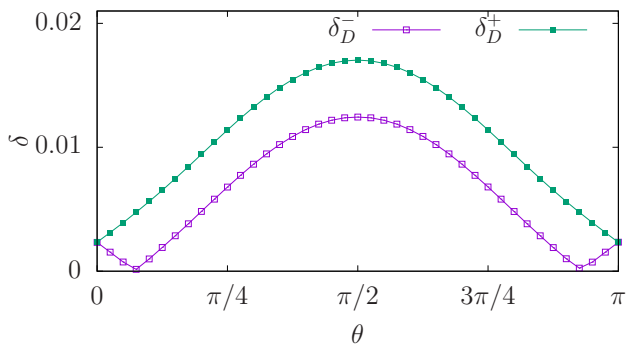}}
\caption{Behavior of the gap at the two Dirac points as a function of the modulation phase $\theta$, for $\omega=1$, $q_{0}=0.5$, $\epsilon/|t_{1}|=1$ (that is, along the dashed line in Fig. \ref{fig:pd}).}
\label{fig:gap}
\end{figure}

In the following, as a representative case we choose $\omega=1$, $q_{0}=0.5$, close to the regime of the experiment in Ref. \cite{jotzu2014}.
The corresponding topological phase diagram is shown in Fig. \ref{fig:pd}. Depending on the values of $\epsilon$ and $\theta$ -- that measure the breaking of inversion and time-reversal symmetries, respectively --, the system can be a topological insulator with Chern number $c=\pm1$ or a normal insulator, $c=0$. 
The boundary between different phases is characterized by the vanishing of the gap at one of the two Dirac points. In case of a formal correspondence between the Haldane model and the effective Hamiltonian $h^{eff}(\bm{k})$ the phase boundaries would be described by an equation of the form 
\begin{equation}
\epsilon=\pm 3\sqrt{3}|t_{1}'|\sin\varphi
\label{eq:gaphaldane2}
\end{equation}
with $t_{1}'$ being a renormalized next-nearest tunneling, and $\varphi$ the corresponding phase (cfr. Eq. (\ref{eq:gaphaldane})). This is shown in Fig. \ref{fig:pd} for $t_{1}'=t_{1}$ and $\varphi=\theta$, as a reference (the fact that it almost captures the actual boundary is just a coincidence, see later on).
The behavior of the gap at the two Dirac points as a function of the modulation phase $\theta$ is shown in Fig. \ref{fig:gap} for $\epsilon/|t_{1}|=1$, corresponding to the horizontal line in Fig. \ref{fig:pd}. 

Remarkably, the modulation of the lattice position allows to simulate the entire topological diagram, contrarily to what happens for the Haldane model in its original formulation, namely in the presence of a microscopic magnetic field. In fact, in the latter case it has been recently shown that only a small fraction of the nominal phase diagram can be accessed, and that the topological insulator phase is suppressed in the deep tight-binding regime \cite{ibanez-azpiroz2015}.

However, from Fig. \ref{fig:bz} we see that the shaking of the lattice not only modifies the gap at the Dirac points, but also affects their position. In general, this is accompanied also by important deformations of the entire energy spectrum, whose actual configuration depends on the phase $\theta$. This is different from the behavior of the original Haldane model,
whose geometry in momentum space is fixed by the symmetries of the system and does not depend on the phase $\varphi$ of the next-nearest tunneling coefficients \cite{haldane1988}. This will become even more clear from the discussion of the structure of the tunneling coefficients, in the following section.

\subsection{Tight binding coefficients}

\begin{figure}[t]
\centerline{\includegraphics[width=0.95\columnwidth]{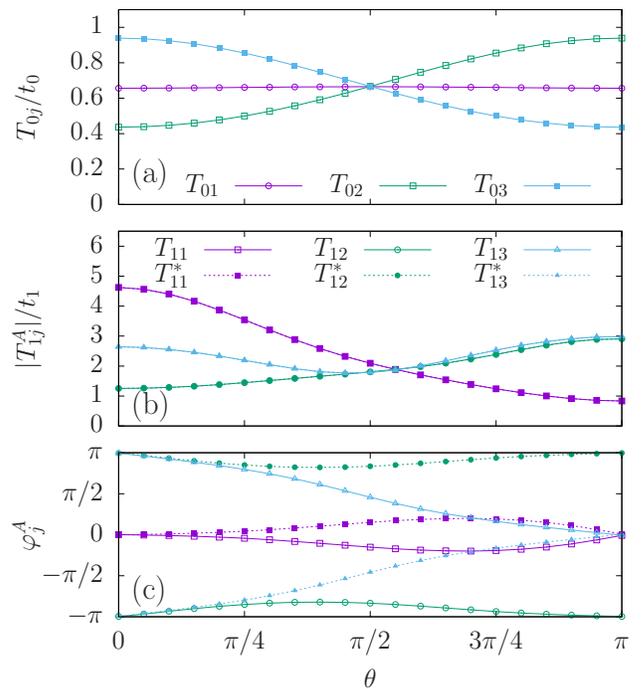}}
\caption{Tunneling coefficients $T_{0j}$ (a) and $T_{1j}^{A}$ (modulus and phase, in (b) and (c) respectively) as a function of the modulation phase $\theta$, for $\omega=1$, $q_{0}=0.5$, $\epsilon=0$. The coefficients $T_{1j}^{B}$ (not shown) display a behaviour similar to that of $T_{1j}^{A}$, flipped with respect to $\theta=\pi/2$. Filled and empty symbols in (b) and (c) refer to complex conjugate pairs (see text).}
\label{fig:tun}
\end{figure}

From the numerical expression of $h_{\nu\nu'}^{eff}(\bm{k})$, one can compute the tunneling coefficients of the corresponding tight binding model on the direct lattice as (see Appendix)
\begin{equation}
T_{\bm{j}-\bm{j}'}^{\nu\nu'}=\frac{\Omega_{\cal B}}{(2\pi)^{2}}\int_{\cal B} d\bm{k} ~e^{i\bm{k}\cdot\bm{R}_{\bm{j}}}h_{\nu\nu'}^{eff}(\bm{k}),
\end{equation}
where we have omitted the $eff$ superscript for ease of notation.
In the present case (see Fig. \ref{fig:honeycomb}) there are three nearest neighbor tunneling coefficients $T_{0j}$, and three complex conjugate pairs of next-nearest tunneling coefficients $T_{1j}^{\nu}$ for each basis point, of type $\nu=A,B$. In principle, even higher order coefficients may arise. 

An example is shown in Fig. \ref{fig:tun}, where we show $T_{0j}$, $|T_{1j}^{A}|$, and the phases $\phi_{j}^{A}$ of the next-nearest coefficients $T_{1j}^{A}$, as a function of the modulation phase $\theta$, for $\omega=1$, $q_{0}=0.5$, $\epsilon=0$. We label them with $j=1,2,3$, where $j=1$ refers to the tunnelings indicated as $t_{0}$ (dotted lines) and $t_{1}$ (pairs of continuous and dashed lines, representing complex conjugate pairs) in Fig. \ref{fig:honeycomb}, the other being ordered counter-clockwise.
This figure shows that the corresponding tight binding model has a very rich structure, and cannot be directly identified with the Haldane Hamiltonian. In fact, while the Haldane model is characterized by a real $t_{0}$, and a complex $t_{1}$ with modulus $|t_{1}|$ and phase $\pm\varphi$ (see Eq. (\ref{eq:t1})), in the present case the number of parameters is threefold: generally, all the parameters of the same class have different values, except for isolated values of the modulation angle (e.g. the three coefficients $T_{0j}$ are degenerate for $\theta=\pi/2$). Only the reality of $T_{0j}$ and the fact that the next-nearest tunnelings $T_{1j}$ come in conjugate pairs is preserved. 

\begin{figure}
\centerline{\includegraphics[width=0.95\columnwidth]{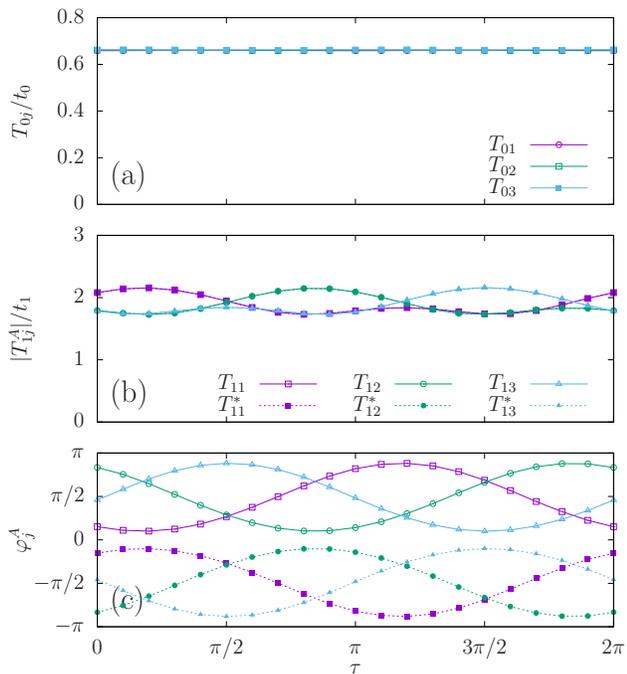}}
\caption{Tunneling coefficients $T_{0j}$ (a) and $T_{1j}^{A}$ (modulus and phase, in (b) and (c) respectively) as a function of the initial driving time $\tau$, for $\theta=\pi/2$, $\omega=1$, $q_{0}=0.5$, $\epsilon=0$. }
\label{fig:tau}
\end{figure}

As anticipated, though the spectrum is gauge invariant (with respect to the choice of the initial driving time $\tau$) \cite{bukov2015}, the value of the tunneling coefficients may be not. As an example, in Fig. \ref{fig:tau} we show their dependence on the initial time $\tau$, for the case of a \textit{circular} path, $\theta=\pi/2$ (the values of the other parameters are $\omega=1$, $q_{0}=0.5$, $\epsilon=0$, as in Fig. \ref{fig:tun}). In particular, this figure shows that the values of $T_{0j}$ are invariant, whereas both $|T_{1j}^{A}|$ and $\phi_{j}^{A}$ show a periodic dependence on $\tau$. Moreover, while the three $T_{0j}$ are degenerate, the rotational symmetry is broken at the level of the $T_{1j}^{A}$, reflecting the fact that different starting points on the circular path (corresponding to different choices of the initial driving time $\tau$) to are not equivalent. This feature goes along with the emergence of the phases $\phi_{j}^{A}$ as a consequence of the non commutativity of the driven Hamiltonian at different times. 

\section{General discussion}
\label{sec:discussion}

\begin{figure}
\includegraphics[width=0.95\columnwidth]{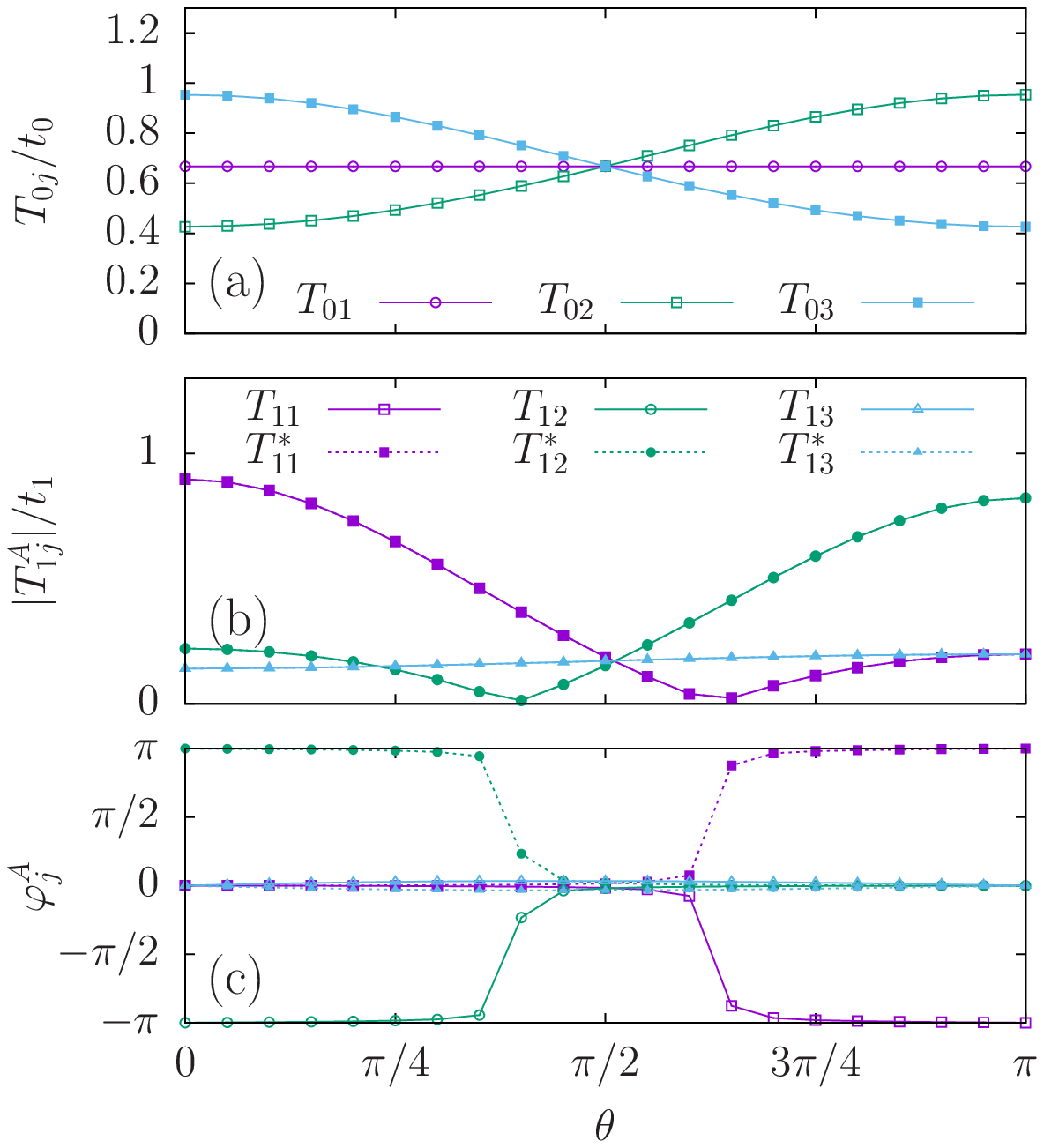}
\includegraphics[width=0.95\columnwidth]{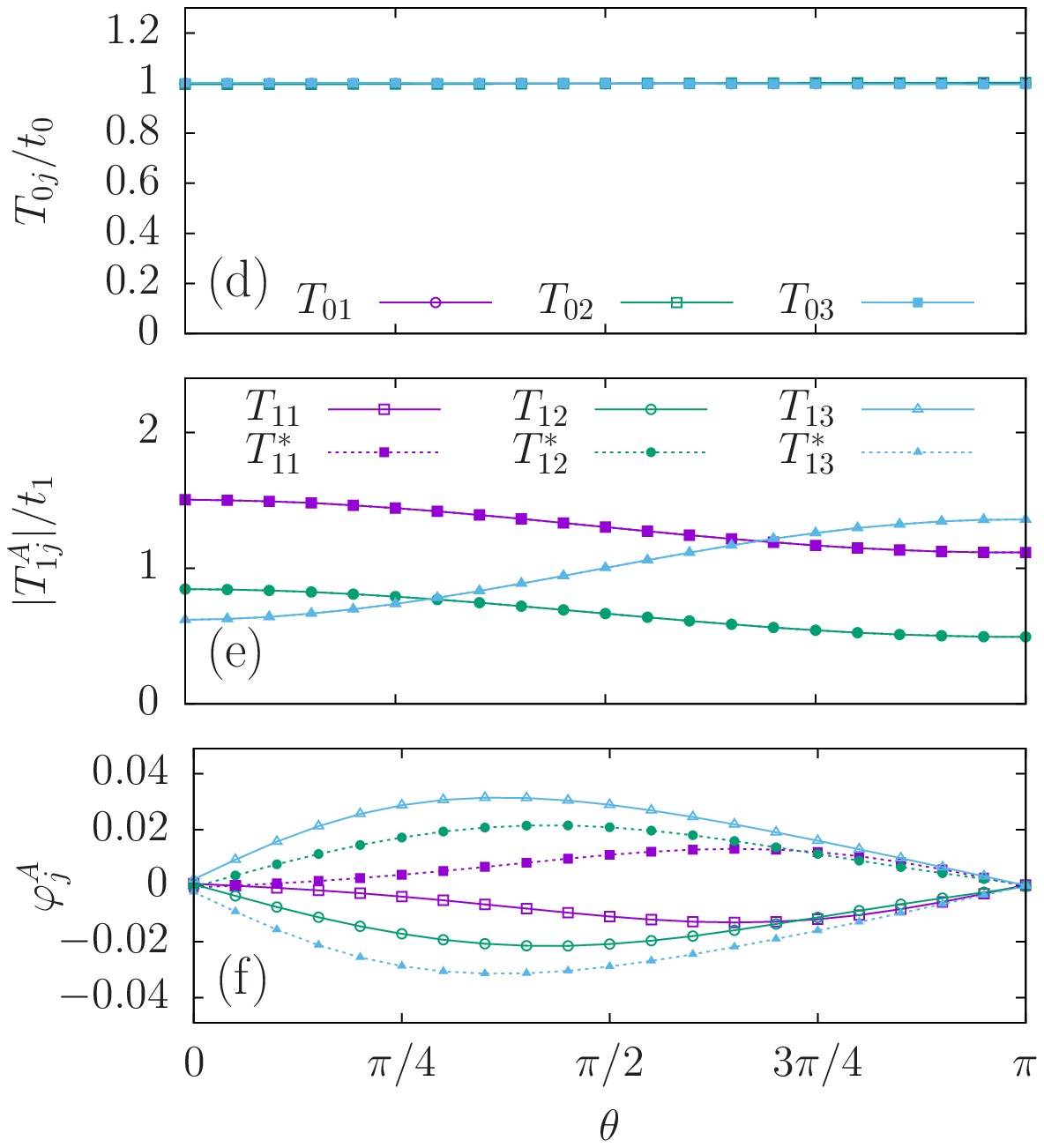}
\caption{Same of Fig. \ref{fig:tun}, here for $\omega=100$ (a-c) and $q_{0}=0.05$ (d-f), all the other parameters being unchanged.}
\label{fig:tun2}
\end{figure}

A systematic analysis of the Floquet Hamiltonian for different values of $\omega$ and $q_{0}$ reveals that the overall behavior discussed in the previous section is very general, and not limited to the specific example considered there. This indicates that indeed there cannot be a formal correspondence between the effective Hamiltonian $h^{eff}(\bm{k})$ (or the corresponding one in the direct lattice) and the Hamiltonian of the Haldane model, although they are characterized by the same topological phases. For example, one cannot identify the phase $\varphi$ of the next-nearest tunneling coefficient of the Haldane model with the modulation phase $\theta$ of the \textit{shaken} honeycomb lattice, though they play the same role in terms of the topological properties of the two systems. In order to make it more clear, we now analyze the behavior of the system in two limiting cases: $\omega\gg1$ (very rapid modulations), and $q_{0}\ll1$ (very small modulation amplitudes).

For $\omega\gg1$, one may expect the effective Hamiltonian to be accurately described by the lowest orders of the Magnus expansion. As discussed in Ref. \cite{jotzu2014,verdeny2015}, this amounts to a renormalization of the tunneling amplitudes (at lowest order), namely of the modulus of $T_{0j}$ and $T_{1j}$, and to the appearance of an imaginary component in the next-nearest terms $T_{1j}$. 
This is consistent with the behavior shown in Fig. \ref{fig:tun2}, especially considering that the phases are close to $0$ or $\pi$. Nevertheless,
even in this limit the tunneling coefficients present a threefold structure and one cannot approximate them with a single set $\{t_{0},|t_{1}|\exp(\pm i\varphi)\}$ as for the Haldane model. 

In the limit of very small modulation amplitudes, $q_{0}\ll1$, the effective Hamiltonian is characterized by a single, unrenormalized, nearest-neighbor tunneling coefficient $T_{0j}=t_{0}$, see Fig. \ref{fig:tun2}d. But still, the fact that the next-nearest tunneling is threefold and that the corresponding phases are very small for any value of the modulation phase $\theta$, does not permit an identification with the Haldane model. We recall that for $q_{0}\equiv0$ the effect of the modulation is vanishing, and one recovers the time-reversal symmetric Hamiltonian in Eq. (\ref{eq:honeytb}).

We also notice that in both limits, $\omega\gg1$ and $q_{0}\ll1$ the topological phase is dramatically squeezed in the vertical direction, as one would expect: in case of a tiny breaking of time reversal symmetry, a tiny breaking of parity is sufficient to bring the system into the normal insulating phase again. In addition, we have not found any limiting case where the boundary between the normal and topological insulator phases can be described by the expression in Eq. (\ref{eq:gaphaldane2}), using for $|t_{1}'|$ some average of the moduli $|T_{1j}|$ of the next-nearest tunneling coefficients, not even assuming (this is not the case) $\varphi\approx\theta$.

\section{Conclusions}
\label{sec:conclusions}

We have discussed the properties of the effective Floquet Hamiltonian for a tight binding model defined on a regular honeycomb lattice, that is periodically modulated in position along an elliptical path as in the recent experiment of Ref. \cite{jotzu2014}. We have found that the effective Hamiltonian of the shaken lattice has a very rich structure and cannot be formally related to the Haldane model. In particular, the effective lattice model is characterized by three inequivalent nearest neighbor and three next-nearest tunneling amplitudes, the latter with different phases $\varphi_{ij}$ that cannot be identified with the phase $\theta$ of the modulation, nor can be directly considered for drawing the topological phase diagram. However, even though the effective model has a richer structure that the Haldane one, the two share the same qualitative behavior of the local structure at the Dirac points, that is dictated by the breaking of parity and time reversal symmetry -- and therefore the same topological phase diagram -- as in the experimental realization of Ref. \cite{jotzu2014}.

\acknowledgments
We thank G. Jotzu for stimulating comments and suggestions. We acknowledge support by the Spanish Ministry of Economy, Industry and Competitiveness and the European Regional Development Fund FEDER through Grant No. FIS2015-67161-P (MINECO/FEDER, UE), and the Basque Government through Grant No. IT986-16.

\appendix
\section{Momentum space transformation}
The Hamiltonian (\ref{eq:honeytb}) can be transformed in momentum space 
(with $\Omega_{\cal B}$ being the volume of the unit cell)
\begin{equation}
\hat{d}_{\nu\bm{k}}=\sqrt{\frac{\Omega_{\cal B}}{(2\pi)^{2}}}\sum_{\bm{j}} ~e^{i\bm{k}\cdot\bm{R}_{\bm{j}}}\hat{c}_{\bm{j}\nu}
\end{equation}
 that transforms any Hamiltonian of the form
\begin{equation}
\hat{H} = \sum_{\nu,\nu'}\sum_{\bm{j,j'}}{\hat{a}}^{\dagger}_{\bm{j}\nu}{\hat{a}}_{\bm{j}'\nu'}T_{\bm{j}-\bm{j}'}^{\nu\nu'}
\end{equation}
into
\begin{equation}
\hat{H}=\sum_{\nu\nu'}\int_{\cal B} d\bm{k}~h_{\nu\nu'}(\bm{k})
\hat{d}_{\nu\bm{k}}^{\dagger}\hat{d}_{\nu'\bm{k}}.
\end{equation}
where
\begin{equation}
h_{\nu\nu'}(\bm{k})=\sum_{\bm{i}}e^{-i{\bm{k}}\cdot{\bm{R}}_{\bm{i}}}
T_{\bm{j}-\bm{j}'}^{\nu\nu'}.
\end{equation}
Given a certain $h_{\nu\nu'}(\bm{k})$, the inverse transformation reads
\begin{equation}
T_{\bm{j}-\bm{j}'}^{\nu\nu'}=\frac{\Omega_{\cal B}}{(2\pi)^{2}}\int_{\cal B} d\bm{k} ~e^{i\bm{k}\cdot\bm{R}_{\bm{j}}}h_{\nu\nu'}(\bm{k}),
\end{equation}
where we have used the summation rule (valid for an infinite lattice)
\begin{equation}
\frac{\Omega_{\cal B}}{(2\pi)^{2}}\sum_{\bm{j}}e^{i{\bm{R}}_{{\bm{j}}}\cdot (\bm{k}'-\bm{k})}=\delta(\bm{k}'-\bm{k}),
\label{eq:deltak}
\end{equation}
The latter formula permits for example to compute the tunneling coefficients of the effective tight-binding Hamiltonian (\ref{eq:heff}).

%

\end{document}